\documentclass[copyright,creativecommons]{eptcs}
\usepackage{preamble}

\begin{document}
\title{Automated Grading of Automata with ACL2s}
\author{Ankit Kumar, Andrew Walter \& Panagiotis Manolios
\institute{Northeastern University}\\
\email{\{ankitk,atwalter,pete\}@ccs.neu.edu}}

\def\authorrunning{Kumar, Walter \& Manolios}
\def\titlerunning{Automated Grading of Automata with {ACL2s}}
\maketitle

\begin{abstract}
  Almost all Computer Science programs require students to take a
  course on the Theory of Computation (ToC) which covers various
  models of computation such as finite automata, push-down automata
  and Turing machines. ToC courses tend to give assignments that
  require paper-and-pencil solutions. Grading such assignments takes
  time, so students typically receive feedback for their solutions
  more than a week after they complete them. We present the Automatic
  Automata Checker (\ag), an open source library that enables one to
  construct executable automata using definitions that mimic those
  found in standard textbooks~\cite{tocbook}. Such constructions are
  easy to reason about using semantic equivalence checks, properties
  and test cases. Instructors can conveniently specify solutions in
  the form of their own constructions. \ag\ can check for semantic
  equivalence between student and instructor solutions and can
  immediately generate actionable feedback, which helps students
  better understand the material. \ag\ can be downloaded and used
  locally by students as well as integrated into Learning Management
  Systems (LMS) like Gradescope to automatically grade student
  submissions and generate feedback. \ag\ is based on the ACL2s
  interactive theorem prover, which provides advanced methods for
  stating, proving and disproving properties. Since feedback is
  automatic, \ag\ can be deployed at scale and integrated into
  massively open online courses.
\end{abstract}

\renewcommand{\arraystretch}{1.5}

\section{Introduction}

In Theory of Computation (ToC) courses, students study models of
computation, including Deterministic Finite Automata (DFAs), Push-Down
Automata (PDAs) and Turing machines (TMs). When constructing automata,
students benefit from getting immediate, automatic feedback. We
present Automatic Automata Checker (\ag)~\cite{a2c}, an open source
library based on the ACL2s theorem
prover~\cite{acl2swebpage,DillingerMVM07,acl2s11}. For the purposes stated
above, \ag\ provides convenient forms for defining, executing, testing
and reasoning about automata.

Automata defined using \ag\ are not only executable, but are also a
formal model of computation that can be reasoned about. Instructors
can use ACL2s functionality to specify and check properties over these
models, or use the full power of the ACL2s theorem prover to implement
custom checks. This advantage is usually missing from visual or XML
based representations, such as those used in Automata Tutor
V3~\cite{automatatutorv3} or JFLAP~\cite{jflap-book}. Furthermore,
access to a powerful theorem prover allows an instructor to test their
constructions, prove properties or even add new theories or
models. \ag\ has built-in support for grading student-submitted
automata and outputting results as JSON, making it easy to integrate
into existing learning management systems. It comes with
out-of-the-box support for Gradescope, a widely-used online grading
system.

When a student submits an automata construction to Gradescope, it is
checked for consistency and semantic equivalence with solutions
provided by the instructor. Students receive immediate feedback, \eg,
if their solution is determined to be incorrect, they are provided
with inputs which their automata incorrectly classify. This is a
significant improvement over existing tools like
JFLAP~\cite{jflap-book} and OpenFLAP~\cite{teaching-formal-languages}
which do not support personalized, high quality feedback. Based on the
feedback they receive, students are then able to update and resubmit
their automata.

\ag\ currently supports 3 models of computation -- DFAs, PDAs and TMs,
in a single package. Input validation, equivalence testing and
property-based testing are supported across all three models. We provide a
uniform testing interface across all supported models of computation,
unlike Automata Tutor V3 or JFLAP, whose level of testing capabilities
depend heavily on the kind of automata being checked.

\ag\ is built on top of ACL2s, which gives access to several useful
features like a powerful macro system to define forms convenient for
declaration of automata (our forms faithfully reflect their
corresponding textbook~\cite{tocbook} descriptions), the
\emph{Defdata}~\cite{defdata} framework to define automata
components as types, theorem proving to prove equivalence of types,
property-based testing for checking equivalence between automata
constructions and to test if such constructions satisfy certain
properties and counter-example generation to produce helpful feedback,
if automata constructions do not satisfy properties. Since \ag\ is
open source, it can be easily extended to support other models or
theories, for example, recursive function theory which has built-in
support in ACL2s.

\textbf{Our contributions.} We make the following contributions with
\ag: (i) an open source library to construct executable models of
DFAs, PDAs and TMs, (ii) ability to reason about constructed automata
using properties and test cases, (iii) instant feedback generation to
guide students towards a correct construction, (iv) ability to run
either locally or on compatible LMS like Gradescope and (v) ability to
extend with more models of computation, with a little experience
working with ACL2s.  To the best of our knowledge, there exists no
other tool with all of these capabilities.

\textbf{Paper Outline.}  Section~\ref{sec:related-work} presents
related work in the field of automated grading of
automata. Section~\ref{sec:acl2s} explains our choice of the ACL2s
theorem prover. Section~\ref{sec:examples} illustrates the kinds of
helpful feedback \ag\ generates on student
submissions. Section~\ref{sec:implimentation} describes implementation
considerations of the system. Section~\ref{sec:system-description}
shows the system architecture and details integration with
Gradescope. Section~\ref{sec:experiences} discusses our experiences
using \ag\ to automatically grade (autograde) assignments and
exams. Section~\ref{sec:limitations} lists some limitations of \ag\
and Section~\ref{sec:conclusion} concludes.

\section{Related Work}
\label{sec:related-work}

Several tools and techniques for automatically grading or providing
feedback for student submitted automata exist. For brevity, we will
describe some of the most commonly used tools, and touch on the
possibility of using specialized algorithms for automata equivalence
checking.

The JFLAP (Java Formal Language and Automata Package) is a
visualization and teaching tool for formal languages. It supports
DFAs, PDAs, and TMs in addition to several other models of computation
and parsing algorithms. JFLAP does not have built-in support for
grading (as it is a tool intended for students to run themselves while
writing up their homework solutions), though several extensions to
JFLAP have been made that attempt to add such
features~\cite{enhancing-jflap, norton2009algorithms}. Both of these
extensions only support DFAs, and the work of Shekhar \etal\ simply
performs bounded checking of words to check equivalence between the
student's submission and the instructor's solution. Neither of these
extensions allow for integration with an external LMS. JFLAP claims to
be open-source, though we could not find any way to acquire the JFLAP
source from its official website.

A more modern incarnation of FLAP called OpenFLAP is a component of
OpenDSA~\cite{opendsa}, an open-source and interactive eTextbook
system. OpenFLAP provides support for auto-graded DFA, PDA, and TM
construction exercises, but student submissions are evaluated solely
on whether they pass concrete test cases provided by the
instructor. OpenDSA can be integrated with the Canvas LMS.

Automata Tutor v3~\cite{automatatutorv3} is a closed-source online
platform that automatically grades and provides feedback on
automata. It has been used to teach thousands of students at over 30
universities, with high reported satisfaction among both students and
instructors~\cite{automatatutorv3}, which highlights the benefits of
automated grading systems in undergraduate level theory of computation
classes. It supports DFAs, PDAs and TMs in addition to several other
models of computation. Instructors can create assignments inside of
Automata Tutor that include automata construction exercises. Students
then use Automata Tutor's interface to graphically construct the
relevant automaton. Automata Tutor automatically grades exercises, and
in the case of DFAs will provide ``descriptive hints'' that aim to
help students understand how their solutions are
incorrect~\cite{dantoni2015how}. For PDAs and TMs, Automata Tutor will
only provide counterexamples (if it can find them). These
counterexamples are generated by testing randomly generated words up
to a configurable length, given certain resource
limits~\cite{automatatutorv3}. A disadvantage of a closed-source
monolithic platform like Automata Tutor v3 is that, it can not be
reliably used when it is experiencing technical problems or is down for
maintenance, as was the case when we last checked on
$12^{th}$ November, $2022$.

Checking equivalence of TMs or (nondeterministic) PDAs is undecidable,
but checking the equivalence of DFAs is decidable and algorithms exist
that can check equivalence and generate a witness (a counterexample to
the equivalence of the two DFAs, \eg, a word that one of the DFAs
accepts but the other rejects) with worst-case runtime complexity
``nearly linear'' in the total number of states in the two
DFAs~\cite{norton2009algorithms}. A complete procedure to check
equivalence of DFAs would be a good future addition to \ag.

\section{ACL2 Sedan}
\label{sec:acl2s}
\ag\ is written in ACL2 Sedan
(ACL2s)~\cite{acl2swebpage,DillingerMVM07,acl2s11}, which is an
extension of A Computational Logic for Applicative Common Lisp
(ACL2)~\cite{acl2webpage,car}. ACL2 is an industrial strength system
for integrated modeling, simulation, and inductive reasoning. It comes
from the Boyer-Moore family of theorem provers and is capable of
reasoning about statements in the first order logic with mathematical
induction. ACL2s extends ACL2 with automation and user friendly
features like an advanced data definition framework Defdata, a
powerful termination analysis based on calling context
graphs~\cite{ccg} and
ordinals~\cite{ManoliosVroon03,ManoliosVroon04,MV05}, a property-based
modeling and reasoning framework for theorem proving, the \emph{cgen}
framework~\cite{harsh-dissertation,cgen,harsh-fmcad,Walter2022} for
generating counter-examples for invalid properties, and support for
systems programming, a relatively new capability that allows one to
build formal-methods-enabled tools that use ACL2s as a key
component and which  has
been used in projects involving gamified verification, education,
proof checking, interfacing with external theorem provers and
security~\cite{interface, walter2021reasoning,
  walter2019gamification}.

\ag\ utilizes all of these features to facilitate writing executable
automata constructions, reasoning about them using properties, and
getting helpful feedback in the form of counter-examples, all in the
same system.

\section{Illustrative Examples}
\label{sec:examples}
Before describing the details of our system, we will walk through a
few illustrative examples that highlight the kinds of feedback that
\ag\ provides to a student. These example problems are also available
on Gradescope as programming assignments, which can be accessed using
instructions provided in~\cite{a2c}.

\subsection{Checking Deterministic Finite Automata}
Consider a homework problem that requires a student to construct a DFA
that can recognize words in $\{0,1\}^*$ consisting of an odd number of
ones. A DFA $M$ is a 5-tuple, $(Q, \Sigma, \delta, q_0, F)$ where $Q$
is a finite set of states, $\Sigma$ is an alphabet,
$\delta : Q \times \Sigma \rightarrow Q$ is a transition function,
$q_0 \in Q$ is a start state and $F \subseteq Q$ is a finite set of
accept states. Suppose the student submits an ACL2s form as shown:

\lstset{language=}
\begin{lstlisting}
  (gen-dfa
   :name     student-dfa
   :states   (e1 e2 o1 o2)
   :alphabet (0)
   :start    e1
   :accept   (o1 o2)
   :transition-fun (((e1 0) . e1) ((e1 2) . o1) ((e2 0) . e2) ((e2 2) . e2)
                    ((o1 0) . o2) ((o1 2) . e2) ((o2 0) . o1) ((o2 2) . e1)))
\end{lstlisting}

\gendfa\ is a macro that generates data definitions for a DFA
construction. ACL2s has a powerful macro system that can be used to
create representations of automata equivalent to those in the
book. Motivated users define their own macros and syntax. The
\gendfa\ form above defines a DFA with the name \sdfa. The order of
components appearing in the form does not matter, \eg, swapping the
order of \st\ and \ac\ yields the same DFA. On reading a \gendfa\
form, \ag\ performs several checks, including ensuring (1) the
name is new, (2) all components of a DFA are provided and (3)
components are well-formed, \eg, that the transition
function is a total function from the appropriate domain to the
appropriate co-domain.

Given the form above, \ag\ reports that \tf\ ($\delta$) is not a
function with domain $ Q \times \Sigma $. This is because
the \ab\ ($\Sigma$) specified in the above form contains the element
\zero, whereas the domain of \codify{:transition-fun} seems to have
been constructed with a different alphabet in mind, namely one which
includes both \zero\ and \two\ as elements. After receiving this
feedback, the student can update \ab\ as follows.

\lstset{language=}
\begin{lstlisting}[belowskip=-0.2 \baselineskip]
  :alphabet       (0 2)
\end{lstlisting}
The updated form passes all of our system's checks, indicating that
the student's submitted form represents a valid DFA. What remains is
to check whether this construction is correct. The specification for a
correct automaton is a solution provided by the instructor:

\lstset{language=}
\begin{lstlisting}[belowskip=-0.2 \baselineskip]
  (gen-dfa
   :name           instructor-dfa
   :states         (even odd)
   :alphabet       (0 1)
   :start          even
   :accept         (odd)
   :transition-fun (((even 0) . even) ((even 1) . odd) ((odd 0) . odd) ((odd 1) . even)))
\end{lstlisting}

Testing for correctness reduces to testing equivalence between the
student's solution \sdfa\ and the instructor's solution \idfa. Part of
checking equivalence of automata includes checking for equivalence of
alphabet. So, in the running example, \ag\ tries to prove the
equivalence of both alphabets, fails and generates the following
feedback:

\lstset{language=}
\begin{lstlisting}[belowskip=-0.2 \baselineskip]
  Incorrect alphabet provided.
\end{lstlisting}

This is because the student's alphabet does not match with the one
provided in the instructor's DFA. The student corrects their alphabet
and updates their transition function as appropriate.

\lstset{language=}
\begin{lstlisting}[belowskip=-0.2 \baselineskip]
  :alphabet (0 1)
  :transition-fun (((e1 0) . e1) ((e1 1) . o1) ((e2 0) . e2) ((e2 1) . e2)
                   ((o1 0) . o2) ((o1 1) . e2) ((o2 0) . o1) ((o2 1) . e1))
\end{lstlisting}

But \sdfa\ is still incorrect, as is pointed out by \ag\ in the
following feedback:
\begin{lstlisting}[belowskip=-0.2 \baselineskip]
  Transition function error. The following words are misclassified:
  ('(0 1 1 1) '(1 1 1 0) '(1 1 1))
\end{lstlisting}

Each list in the feedback represents a word that is misclassified by
\sdfa. With this feedback, the student learns something new about their
construction and can run their DFA on the generated words as shown:

\lstset{language=}
\begin{lstlisting}[belowskip=-0.2 \baselineskip]
  (run-dfa student-dfa '(0 1 1 1))
\end{lstlisting}
The \codify{run-dfa} form runs \sdfa\ on the input word
\codify{'(0 1 1 1)} generating the following output:

\lstset{language=}
\begin{lstlisting}[belowskip=-0.2 \baselineskip]
  E2
\end{lstlisting}

This helps the student realize that any transition from \et\ leads
back to \et. The student corrects this mistake as shown:

\lstset{language=}
\begin{lstlisting}[belowskip=-0.2 \baselineskip]
  :transition-fun (((e1 0) . e1) ((e1 1) . o1) ((e2 0) . e2) ((e2 1) . o2)
 	           ((o1 0) . o2) ((o1 1) . e2) ((o2 0) . o1) ((o2 1) . e1))
\end{lstlisting}

The updated solution is finally accepted by the autograder, which outputs:

\lstset{language=}
\begin{lstlisting}[belowskip=-0.2 \baselineskip]
  student-dfa is correct.
\end{lstlisting}

\subsection{Checking Push Down Automata}
Consider another homework problem, one that requires a student to
construct a a PDA that can recognize the language
$\{0^n1^n \ | \ n \geq 0 \}$. Recall that a PDA is a 6-tuple,
$(Q, \Sigma, \Gamma, \delta, q_0, F)$ where $Q$ is a finite set of states,
$\Sigma$ is the input alphabet, $\Gamma$ is the stack alphabet,
$\delta : Q \times (\Sigma \cup \{\epsilon\}) \times (\Gamma \cup \{\epsilon\}) \rightarrow \mathcal{P}( Q \times (\Gamma\cup \{\epsilon\}))$ is
a transition function, $q_0 \in Q$ is a start state and
$F \subseteq Q$ is a finite set of accept states.

Consider a student's submission for this problem:

\lstset{language=}
\begin{lstlisting}
  (gen-pda
   :name student-pda
   :states (q1 q2 q3)
   :alphabet (0 1)
   :stack-alphabet (0 z)
   :start-state q1
   :accept-states (q3)
   :transition-fun (((q1 0 :e) . ((q1 0)))
                    ((q1 1  0) . ((q2 :e)))
                    ((q2 1  0) . ((q2 :e)))
                    ((q2 :e z) . ((q3 :e)))))
\end{lstlisting}

The \genpda\ form shown above defines a PDA with the name
\spda. Similar to the \gendfa\ form described earlier, the order of
components in this form does not matter. We use \codify{:e} to
represent the empty word $\epsilon$. We could have chosen to use the unicode
character for $\epsilon$ as well, but we did not want students to deal with
potential issues due to lack of unicode support. Requiring only ASCII
characters as input allows \ag\ to support online courses where
students might use a plethora of development environments to write
their solutions. Most of the checks related to validity of states,
input alphabet and stack alphabet are similar to those for DFAs. Even
though we do not require transitions from every possible tuple of
state, alphabet and stack symbols (we assume empty set by default), we
do include a check for the presence of a transition from the start
state on the empty word \codify{:e}. This is required to add a base
stack symbol. The submission shown above fails this test and hence the
student receives the following feedback:

\lstset{language=}
\begin{lstlisting}[belowskip=-0.2 \baselineskip]
  Starting transition from (Q1 :e :e) missing from the transition function.
\end{lstlisting}

This is fixed by adding a new start state q0 and updating the transition
function as shown:
\begin{lstlisting}
  (gen-pda
   :name student-pda
   :states (q0 q1 q2 q3)
   :alphabet (0 1)
   :stack-alphabet (0 z)
   :start-state q0
   :accept-states (q3)
   :transition-fun (((q0 :e :e) . ((q1 z)))
                    ((q1 0  :e) . ((q1 0)))
                    ((q1 1   0) . ((q2 :e)))
                    ((q2 1   0) . ((q2 :e)))
                    ((q2 :e  z) . ((q3 :e)))))
\end{lstlisting}

The updated solution passes all checks for a valid PDA. It is now time
to check whether it matches the specifications of the problem. After
checking if the submitted PDA is correct, the autograder reports

\lstset{language=}
\begin{lstlisting}[belowskip=-0.2 \baselineskip]
  Transition function error. The following words were misclassified :
  (:e)
\end{lstlisting}

This feedback suggests that \spda\ does not correctly classify the
input word $\epsilon$. Indeed, this word should be accepted, but is not
accepted by \spda. This can be fixed either by modifying the set of
accept states:
\begin{lstlisting}
  :accept-states (q0 q3)
\end{lstlisting}
or by modifying the transition function to allow an $\epsilon$ transition
from the start state to the accept state:
\begin{lstlisting}
  :transition-fun (((q0 :e :e) . ((q1 z) (q3 :e)))
                   ((q1 0  :e) . ((q1 0)))
                   ((q1 1   0) . ((q2 :e)))
                   ((q2 1   0) . ((q2 :e)))
                   ((q2 :e  z) . ((q3 :e))))
\end{lstlisting}

This updated solution is finally accepted upon submission.

\lstset{language=}
\begin{lstlisting}[belowskip=-0.2 \baselineskip]
 student-pda is correct.
\end{lstlisting}

\subsection{Checking TMs}
Consider a problem where a student is required to submit a TM that
flips 0s to 1s and vice-versa on its tape. A TM is defined as a
7-tuple $(Q,\Sigma,\Gamma,\delta,q_0,q_{\mbox{accept}},q_{\mbox{reject}})$ where $Q$ is a finite
set of states, $\Sigma$ is the input alphabet, $\Gamma$ is the tape alphabet,
$\delta : Q \times \Gamma \rightarrow Q \times \Gamma \times \{L, R\}$ is the transition function,
$q_0 \in Q$ is the start state, $q_{\mbox{accept}}$ is the accept state and
$q_{\mbox{reject}}$ is the reject state. We require that the execution of the
TM on the given input $w$ end with the head at the end of the output
on its tape. We ignore any symbols occurring to the right of the head
after the end of execution. We also trim all blank symbols occurring
between the end of the output and the head.

Suppose that a student submits the following TM for the above problem:

\begin{lstlisting}
  (gen-tm
   :name student-tm
   :states (q0 q1 q2 q3)
   :alphabet (0 1)
   :tape-alphabet (0 1)
   :start-state q0
   :accept-state q1
   :reject-state q2
   :transition-fun (((q0 1) . (q0 0 R))
                    ((q0 0) . (q0 1 L))
                    ((q0 nil) . (q3 nil R))
                    ((q3 nil) . (q1 nil L))))
\end{lstlisting}

The \gentm\ form shown above defines a TM with the name \stm. Similar
to the previously seen \codify{gen-} forms, the order of components in
this form does not matter. After submitting this form, the following
feedback is produced by the autograder:

\lstset{language=}
\begin{lstlisting}[belowskip=-0.2 \baselineskip]
  Blank tape symbol nil missing from tape-alphabet.
\end{lstlisting}

The blank tape symbol is represented using the keyword
\codify{nil}. The \codify{tape-alphabet} component of \codify{gen-tm}
is required to include the blank tape symbol \codify{nil} and the
input alphabet has to be a subset of the tape alphabet. This is fixed
by modifying \codify{tape-alphabet} :

\begin{lstlisting}
  :tape-alphabet (0 1 nil)
\end{lstlisting}

After editing and resubmitting, the autograder gives more feedback:

\lstset{language=}
\begin{lstlisting}[belowskip=-0.2 \baselineskip]
  Incorrect output produced when running submitted TM on the following words :
  ('(1 0) '(0) '(0 1 1))
\end{lstlisting}

Running their TM on words provided in the feedback, the student
realizes a mistake in their transition function, and corrects it as
shown:

\begin{lstlisting}
  :transition-fun (((q0 1) . (q0 0 R))
                   ((q0 0) . (q0 1 R))
                   ((q0 nil) . (q3 nil R))
                   ((q3 nil) . (q1 nil L)))
\end{lstlisting}

The updated solution is finally accepted by the autograder.

\section{Implementation}
\label{sec:implimentation}
As already mentioned, our tool is based on the ACL2s theorem
prover~\cite{DillingerMVM07,acl2swebpage}, an extension of
ACL2~\cite{car,acl2webpage} which consists of a programming language,
a logic for the language and an interactive theorem prover.

\subsection{Solution format}
\ag\ provides a well defined input format for specifying automata. A
solution file in this format consists of one or more \genx\ forms
where $x$ may be one of \codify{dfa}, \codify{pda} or
\codify{tm}. Each such form provides a declarative description of its
corresponding automaton. The description is faithful to its textbook
definition~\cite{tocbook}. Hence the input format can be naturally
explained and is easy to understand.

\subsection{Validating automata}
\ag\ validates a \genx\ form by checking whether all of the
following conditions hold:
\begin{itemize}
\item all required components of the automaton are provided,
\item the start state is one of the given states,
\item the set of accept states is a subset of the set of states,
\item the domain of the transition function is of the right type,
\item the co-domain of the transition function is of the right type,
\item additional model-specific checks for each of PDAs and TMs, \eg,
 the blank tape symbol does not appear in the alphabet of a TM, but
 should appear in the tape-alphabet.
\end{itemize}

\subsection{Checking correctness}
\ag\ uses the defdata data definition framework to convert an
automaton description into corresponding definitions of states,
alphabets, transition-functions and functions which make it executable.

For property-based testing, instructors can either use
\codify{property} forms (for testing as well as theorem proving) or
\codify{test?} forms (meant solely for testing). Property-based
testing depends on the cgen framework for generating counter-examples
to invalid properties. \ag\ makes use of the
\codify{interface}~\cite{interface} library to query ACL2s with
\codify{test?} forms to test equivalences and extract
counter-examples.

\textbf{Cgen framework:} property-based testing using ACL2s'
\codify{test?} forms depends on cgen, a counter-example generation
framework which combines theorem proving with testing in a synergistic
way. The framework is quite mature and has been used extensively in a
number of projects, including industrial
projects.~\cite{moitra2018towards, mcmillan2019increasing} It provides
numerous configuration options and is based on a number of algorithms
and ideas that are described in related work.  In brief, it uses a
collection of algorithms and the full power of the theorem prover to
simplify conjectures and to decompose them into subgoals, \eg, it may
prove that a conjecture holds in certain infinite regions of the
state space, thereby removing these regions from further
consideration. It also uses a collection of testing methods that are
performed at key parts of the theorem proving process. The testing is
integrated with the underlying ACL2s type system to generate random
elements of given types, optionally satisfying certain constraints.
The counter-example generation framework can even improve the ability
of ACL2s to prove theorems, \eg, it may reveal that a generalization
step falsifies a given conjecture, at which point it forces the
theorem prover to backtrack, which avoids certain failure and may lead
to a proof. The number of tests generated for property-based testing
is configurable. For use in our undergraduate-level ToC class, we used
the default number of test cases for \codify{test?} forms, which is
1000, and are not aware of any false-positives.

\textbf{Testing equivalence:} To test if two automata are
 equivalent, we first check if their alphabets match:

\lstset{language=}
\begin{lstlisting}[belowskip=-0.2 \baselineskip]
  (defdata-equal instructor-x-alphabet student-x-alphabet)
\end{lstlisting}

Here \codify{instructor-x-alphabet} and \codify{student-x-alphabet}
are definitions generated by either of \gendfa, \genpda\ or \gentm,
(depending on \codify{x} being either of \codify{dfa}, \codify{pda} or
\codify{tm}) for the alphabet of the instructor and the student
automata respectively. \codify{defdata-equal} checks whether two data
types are identical. If not, the defdata framework~\cite{defdata} will
generate a value that is in one of the data types and not in the
other. In this case, if the student and instructor alphabets differ,
defdata will return a symbol that is in one alphabet but not the
other. To complete the check of equivalence, we check the following
property.

\lstset{language=}
\begin{lstlisting}[belowskip=-0.2 \baselineskip]
  (test?
    (=> (instructor-x-wordp w)
        (== (accept-x w *student-x*)
            (accept-x w *instructor-x*))))
\end{lstlisting}
where \codify{x} is one of \codify{dfa, pda} or \codify{tm}.

Note that whenever a \codify{gen-x} form successfully validates and
generates data definitions for each component of a given construction
named \codify{student-x}, behind the scenes, it generates a constant
\studentx\ to refer to a list consisting of the given components. The
(\acceptx\ \studentx\ \codify{w}) and (\acceptx\ \instructorx\
\codify{w}) forms check whether the student's automaton and the
instructor's automaton accept the word \codify{w} of type
\codify{instructor-x-word} respectively. Any counter-example generated
by this \codify{test?} form indicates a word accepted by exactly one
of the automata. The number of steps PDAs and TMs are allowed to run
for on an input is bounded and can be configured by the
instructor. In case of TMs, we utilize another test for equivalence,
one that checks whether the two TMs agree on their output.

\lstset{language=}
\begin{lstlisting}[belowskip=-0.2 \baselineskip]
  (test?
     (=> (instructor-tm-wordp w)
         (== (remove-final-nils (left-of-head (run-tm w *student-tm*)))
             (remove-final-nils (left-of-head (run-tm w *instructor-tm*))))))
\end{lstlisting}

where \codify{left-of-head} of the output of \codify{run-tm} is the
part of the tape to the left of the head, where we expect our output
to reside. \codify{remove-final-nils} removes blank symbols occurring
between the end of the output and the head.

Automata constructions are executed using functions, whose definitions depend on
the automaton \codify{x}:

\begin{itemize}
\item Given a initial state $q_0$, a transition function $\delta$ and a
  word $w$, \codify{run-dfa} returns a state $s \in Q$. Starting with
  $q_0$ and the first letter in $w$, it queries $\delta$ to get a new
  state. This process is repeated until all letters in $w$ are
  exhausted and $s$ is reached.

  Since $\delta$ is complete (all pairs in $Q \times \Sigma$ belong to
  the domain of $\delta$), and since the length of $w$ is finite,
  \codify{run-dfa} is terminating.

\item
  Consider a tuple $t \in Q \times \Gamma^{*} \times \Sigma^{*}$ denoting a
  state reached, contents in a stack and a suffix of the input
  word left to be consumed, respectively. We call this an \etp.

  \codify{run-pda} is a function that accepts a bound (a natural
  number), a PDA and a set of \etp s, and returns a boolean: t or nil.
  The execution trace of \codify{run-pda} is a tree such that each of
  its nodes is an \etp. The root of this tree is $(q_0,\epsilon,w)$ \ie, a
  tuple consisting of the start state, an empty stack and the input
  word. At each step of the execution, new leaves of the tree may be
  generated. Recall that
  $\delta: Q \times (\Sigma \cup \{\epsilon\}) \times (\Gamma \cup \{\epsilon\}) \rightarrow \mathcal{P}( Q \times (\Gamma\cup
  \{\epsilon\}))$. For a leaf $l = (q_l,(t \ldots),(c \ldots))$ where
  $q_l \in Q$, $t$ is the top of its stack and $c$ is the first letter
  in the rest of the word, we have the following tuples :
  $(q_l,t,c),(q_l,\epsilon,c), (q_l,t,\epsilon)$ and
  $(q_l,\epsilon,\epsilon)$. If any of these tuples exists in the domain of
  $\delta$, we say that leaf $l$ is \emph{active} and the children of
  $l$ can be generated using $\delta$. In case none of these tuples exist
  in the domain of $\delta$, $l$ remains a leaf node.

  An \etp\ is accepted when it is of the form $(q_f,\ldots,\epsilon)$ where
  $q_f \in F$. \codify{run-pda} is executed until either an acceptable
  \etp\ is generated (in which case it returns t) or when all active
  leaves of the tree are at a depth greater than $n$, in which case it
  returns nil.

  The inherent non-determinism of PDAs coupled with the possibility of
  making $\epsilon-$transitions does not allow guaranteed termination of
  \codify{run-pda}. Hence, the execution of \codify{run-pda} needs to
  be bounded by a positive integer $n$, the maximum depth of an
  execution tree decided by the instructor.

\item Given a TM
  $(Q,\Sigma,\Gamma,\delta,q_0,q_{\mbox{accept}},q_{\mbox{reject}})$ and a word
  $w$, \codify{run-tm} returns a tuple
  $(Q \times \Gamma^{*} \times \Gamma^{*})$ whose first element is the current state of
  the TM and the second and third elements are the contents of the
  tape to the left and to the right of the TM's head,
  respectively. Note that tape symbols to the left of the head are
  stored in reverse. The head of a TM moving left on its tape is
  simulated by removing the first element from the tape on the left of
  head and attaching it at the start of the tape on the right of
  head.

  Since the halting problem is undecidable for a Turing machine,
  we need to set a maximum number of steps the TM can run for before
  stopping it.
\end{itemize}

\textbf{Testing properties:} Since checking the equivalence of PDAs
and TMs is undecidable, it is useful for the instructor to have a
collection of properties to check a students' constructions. For
example, in context of our running DFA example,

\lstset{language=}
\begin{lstlisting}[belowskip=-0.2 \baselineskip]
  (property no-odd1s-in-ww (w :instructor-dfa-word)
    :proofs? nil
    (! (accept-dfa *student-dfa* (append w w))))
\end{lstlisting}

checks that *\sdfa*\ does not accept words of the form \codify{ww}. This
is because such words can not have an odd number of \codify{1}s. Such
properties represent an infinite class of test cases a submission can
be evaluated against. Of course, ground expressions such as unit
tests can also be defined. Property forms in ACL2s can be
configured to perform as much testing as one requires. In our example,
we turn off theorem proving locally for this property using
\codify{:proofs? nil} to focus exclusively on testing in hopes of
finding counter-examples. It is possible to disable theorem proving
globally, for all properties as well, as shown:

\begin{lstlisting}[belowskip=-0.2 \baselineskip]
  (set-acl2s-property-table-testing? t)
  (set-acl2s-property-table-proofs? nil)
\end{lstlisting}

Now that theorem proving has been disabled and testing enabled
globally for property forms, we have

\lstset{language=}
\begin{lstlisting}[belowskip=-0.2 \baselineskip]
  (property accept-w->accept-0w1 (w :instructor-pda-word)
    (=> (accept-pda *student-pda* w)
        (accept-pda *student-pda* (app '(0) w '(1)))))
\end{lstlisting}
  
which checks that if a word $w$ is accepted by *\spda*, so is
$0w1$.

Finally, for the TM example, we could use the following property
to test a student submission, after it has been validated:
\lstset{language=}
\begin{lstlisting}[belowskip=-0.2 \baselineskip]
  (property involution (w :student-tm-word)
    (== (remove-final-nils (left-of-head (run-tm
          (remove-final-nils (left-of-head (run-tm w *student-tm*))))))
        w))
\end{lstlisting}

which checks that if running *\stm*\ on the input word $w$ produces $w'$
in which the 0s and 1s are flipped, then running *\stm*\ again on $w'$
should produce $w$ as output.

Notice that each property is specified programmatically in the ACL2s
language in terms of executable automata, making the process of
creating tests effortless.

\textbf{Unit testing:} An instructor can also make use of
\codify{check=} forms to check their own construction for specific test
cases. For example, from our running examples,

\begin{lstlisting}[belowskip=-0.2 \baselineskip]
  (check= (accept-pda *instructor-pda* '(0 0 0 1 1 1)) t)
\end{lstlisting}
checks whether \codify{*instructor-pda*} accepts \codify{'(0 0 0 1 1
  1)}. Similarly, 
\begin{lstlisting}[belowskip=-0.2 \baselineskip]
  (check= (remove-final-nils (left-of-head (run-tm '(1 0 1 1 1 0 1 0) *instructor-tm*)))
     '(0 1 0 0 0 1 0 1))
\end{lstlisting}
checks whether running \codify{*instructor-tm*} on input \codify{'(1 0
  1 1 1 0 1 0 1 0)} flips 0s and 1s.

\section{System Description}
\label{sec:system-description}
\ag\ can be packaged as an executable and deployed on any online
grading platform that can run the executable on student submissions
and can use the output generated to grade assignments.

\subsection{ACL2s and external tools}
\begin{wrapfigure}{r}{5.5cm}
  \vspace{-1cm}
  \centering
  \includegraphics[width=0.28\textwidth]{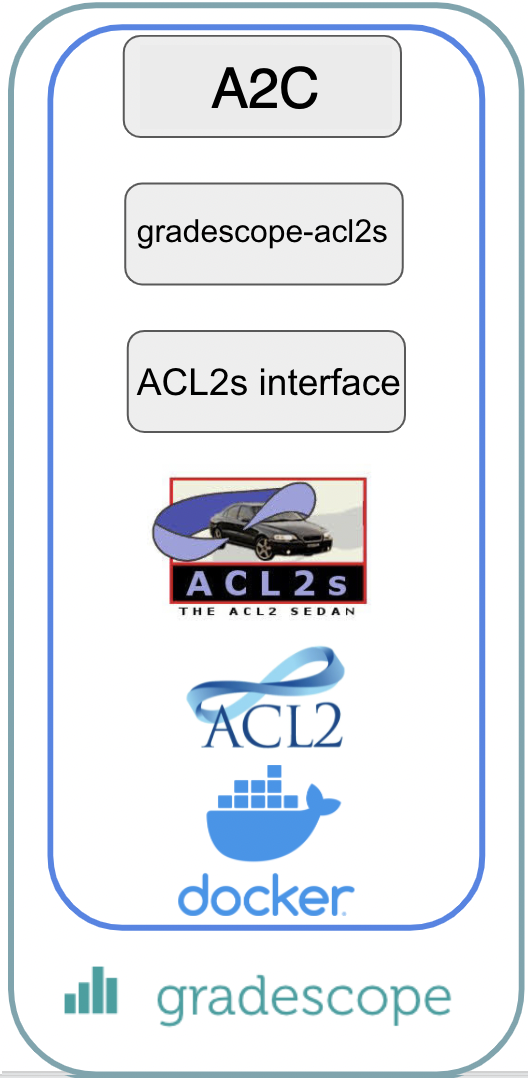}
  \label{techstack}
  \caption{Technical stack for autograding ToC assignments}
  \vspace{-1.5cm}
\end{wrapfigure}
As previously discussed in Section~\ref{sec:acl2s}, ACL2s extends the
ACL2 theorem prover with several automation and user-friendly
features. ACL2 provides a full-featured programming language, so it is
possible to implement and integrate additional functionality inside of
ACL2 without needing to modify or understand its internals. ACL2 is
built on top of Common Lisp, and is designed in such a way that one
can write Common Lisp code that interacts with ACL2 code and vice
versa. This means that it is also relatively easy to integrate
external tools with ACL2.

ACL2s provides a feature that allows a user to save the state of a
running ACL2s process as an executable file. We use this feature to
generate an executable for each assignment that an instructor would
like to autograde. The state of such an executable includes libraries
that we wrote that provide the forms required for testing student
constructions and integrating with Gradescope in addition to the
instructor's version of any automata being graded.

\subsection{Libraries}
Our Gradescope-based automatic grader for the ToC course uses the
following libraries:
\begin{itemize}
\item \codify{\ag}: forms needed to define, validate, run and check
  equivalence between user-provided automata
\item \codify{gradescope-acl2s} : interacts with Gradescope by
  generating JSON files consisting of scores and feedback for
  autograded submissions.
\item \codify{interface} : provides an interface with ACL2s' theorem
  proving and counterexample generation functionality.
\end{itemize}

These libraries are designed to be reusable in other contexts besides
\ag. Students can use the \ag\ library locally to inspect their automata
before submitting. For example, a student who is so inclined can
define an automata and then use ACL2s to check both concrete test
cases and properties that they believe their automata should
satisfy. Instructors may also find this useful to check their work
while developing solution automata. Defining properties requires some
experience working with the ACL2s theorem proving system. Freshman
computer science students generally learn ACL2s as part of
Northeastern University's Logic and Computation class.

\subsection{Gradescope Integration}
Gradescope is a LMS used at several universities. It provides
autograding functionality, and can be configured to allow students to
submit the same assignment multiple times and review autograder
feedback after each submission.

A Gradescope autograder is a zip-archive consisting of code that
accepts student submissions and generates JSON files consisting of
points received and feedback for each autograded solution.  Student
submissions are run in Docker containers, which are setup according to
specifications provided in the autograder zip-archive.  Since
Gradescope Docker containers do not support ACL2s by default, we use a
custom Docker image~\cite{acl2simage}. In most cases, an instructor
can simply use one of our provided examples, just updating the
problems and solutions as necessary, to publish an autograded
assignment.

Our autograder can be adapted to support any LMS that supports
autograding using a Docker image or an executable. To do this, one
would need to write a library analogous to \codify{gradescope-acl2s}
that handles whatever I/O is appropriate for that LMS.

\section{Experiences and Observations}
\label{sec:experiences}
We deployed \ag\ on Gradescope for autograding assignments in a
undergraduate-level ToC class taught at Northeastern University,
consisting of about 50 students. Students were introduced to the input
format for specifying each automata before the corresponding
assignment was released.

\subsection{Ease of use}
Setting up autograded assignments using \ag\ was easy. Solutions were
easy to specify, since they were simply the instructor's solution. A
portion of the final exam consisted of autograded problems. For
example, a question similar to : ``Write a TM to insert a 1 in the
input tape after the third occurrence of symbol \# from the left'' was
asked in the final exam. TMs submitted for this problem were fairly
complex (\ie\ they had several states and transitions). Autograding
helped us save about 16 hours of TA time grading this problem. In
addition to this, usually one in thirty students would have sent
regrade requests, which would have taken another 2 hours. Hence it was
useful for the students as well.

\subsection{Anecdotal evidence}
We observed that in comparison to manually graded assignments: (1)
autograded assignments had a significantly higher number of
resubmissions, (2) students received higher grades in the autograded
assignments and (3) student feedback regarding the autograding was
overwhelmingly positive due to the immediate feedback which allowed
students to find trivial errors and helped them better understand the
course material.  On average, more than 95\% of students got full
credit on autograded problems, whereas less than 20\% got full credit
on manually graded problems.
\section{Limitations}
\label{sec:limitations}
Our approach has few limitations, which we discuss here.
Firstly, our tool does not provide a graphical interface for
constructing or inspecting automata. Although the format we use to
submit automata is similar to that used in Sipser's classic
textbook~\cite{tocbook}, students may still need to learn how to read
and write S-expressions to write automata descriptions. This was not
an issue in our class since most students had already taken the
``Logic and Computation'' class, where they were introduced to
programming in ACL2s. However, in case they do not understand
S-expressions, students may require some additional training before
they can use our tool.

Adding visualizations of automata to our tool would be relatively
straightforward - since our tool is written in ACL2s, it is easy to
develop code that will produce output suitable for visualization
tools. Adding a UI for constructing automata may be somewhat more
difficult and will not integrate with Gradescope. 

Though \ag\ does not require its users to have any ACL2s experience,
some functionality does require ACL2s experience. In particular,
writing an extension of our tool requires the user to write ACL2s
functions. Users may also need to understand ACL2s to some extent to
make use of the property system.

Additionally, even in cases where the automata equivalence problem is
decidable, the method we use to check for automata equivalence is not
complete, which is a result of having a uniform testing interface
across all supported models of computation. Put another way, our tool
may return false positives, \eg, it may report that two automata are
equivalent when they are not. While using our tool in a Theory of
Computation course, we performed spot checks of our tool's results and
did not find any issues. It may be the case that the mistakes that
students make tend to be easy to exploit -- that is, they are easy to
find counterexamples. We expect that more complex automata and
automata that differ in their classification of a small number of
words would be more prone to false positives. To mitigate the issue of
incompleteness, an instructor using our tool can provide both test
cases and properties to check tricky errors that the automata
equivalence checker might miss. If they have some ACL2s expertise, the
instructor can use a variety of techniques to improve ACL2s' ability
to find witnesses to non-equivalence. However, adding more properties
should be sufficient in most cases.

Overall, we believe that trading off incompleteness for ease of use is
worthwhile, as this allows instructors and TAs to spend more time
working with students on the truly interesting parts of the course.

\section{Conclusion and Future work}
\label{sec:conclusion}
We present \ag, a library based on the ACL2s theorem prover to check
and provide automatic feedback to students about automata they
construct. Such constructions are executable and generate a formal
model to reason about in the context of a theorem prover, by checking
for equivalence with the instructor's construction, testing properties
and test cases. It generates immediate and helpful feedback. \ag\ can
either be run locally or on a compatible LMS. It has been integrated
into Gradescope to grade assignments and exams. All of these abilities
make \ag\ ideal for use in massively open online courses, where
instructor-student ratio is generally too low to allow highly
available, high quality and personalized feedback by instructors.  We
believe that a similar approach of theorem prover based feedback
generation can be used for non-programmable assignments in other
subjects like logic and discrete math.

In the future, we would like to add the capability to generate
visualizations of automata in order to make learning more intuitive
and effective for students. We would also like to add decision
procedures for decidable problems, like checking equivalence of
DFAs. However, this is an example of an extension which can be
implemented by end users as well, due to \ag\ being open source.

\subsection*{Acknowledgements}
We sincerely thank Mirek Riedewald and Jason Hemann, the instructors
for the Theory of Computation course at Northeastern University in
Fall 2020 and Fall 2021 respectively.

\nocite{*}
\bibliographystyle{eptcs}
\bibliography{refs.bib}
\end{document}